\documentclass{elsart}
\usepackage{graphicx}
\begin{document}

\begin{frontmatter}

\title{Self-consistent theory of intrinsic localized modes:
application to monatomic chain}
\author[Tartu]{V. Hizhnyakov},
\author[Tartu]{A. Shelkan},
\author[Tallinn]{M. Klopov}
\address[Tartu]{Institute of Physics, University of Tartu, Riia 142,
51014 Tartu, Estonia}
\address[Tallinn]{Institute of Physics, Tallinn University of Technology,
Ehitajate 5, 19086 Tallinn, Estonia}

\date{\today}

\begin{abstract}

A theory of intrinsic localized modes (ILMs) in anharmonic lattices
is developed, which allows one to reduce the original nonlinear
problem to a linear problem of small variations of the mode. This
enables us to apply the Lifshitz method of the perturbed phonon
dynamics for the calculations of ILMs. In order to check the theory,
the ILMs in monatomic chain are considered. A comparison of the
results with the corresponding molecular dynamics calculations shows
an excellent agreement.

\noindent PACS: 63.20.Pw,\,\, 63.10.+a,\,\, 63.22.+m

\end{abstract}

\begin{keyword}
Lattice vibrations; Anharmonicity; Intrinsic localized mode.
\end{keyword}

\end{frontmatter}

\section{Introduction}
It is already a well-known fact that localized vibrational
excitations can exist in perfect anharmonic lattices
\cite{Ovchinnik,kosevkov,dolg,sivtak,Page,KisBick,SievPage,flach,LaiSi,christ}.
Such excitations are called intrinsic localized
modes (ILMs) \cite{sivtak}, discrete breathers \cite{flach} or
discrete solitons \cite{christ}. The main approach, which has been
used for the study of these vibrational excitations so far, is
molecular dynamics (MD) simulations, which are based on the
numerical integration of the classical equations of motion in a
lattices with a finite number of the degrees of freedom. The
approach is rather efficient in the case of simple 1D lattices (see,
e.g.
\cite{Aoki,Swanson,Schwartz,TrMaOr,Bind,Craig,Fleisch,FrBW,Bishop,MSato,FrBWal,FrBWall}),
but requires very lengthy computations in the case of 2D and 3D
lattices due to the rapid growth of the number of numerical
operations with the increase of the number of the vibrational
degrees of freedom (see \cite{KisBS,Schrod,KisSiev}). Therefore, it
is of interest to develop other methods which would allow one, when
studying ILMs, to reduce the amount of numerical computations.

Below we are developing a mean-field-type theory of ILMs which
allows one to perform calculations for macroscopically large
lattices of arbitrary dimensions. The theory is based on the
consideration of small variations of the ILM amplitude. The
equations for these variations include characteristics of the ILM.
This allows us to reduce the nonlinear problem of the ILM to a
linear problem of the perturbed phonon dynamics; the perturbation
comes from the ILM and it is determined self-consistently. To
describe the effect of the perturbation we apply the Greens'
function method  of the perturbed local dynamics of phonons (the
Lifshitz method), allowing us to calculate the ILM. As compared to
the method \cite{sivtak}, which also uses the Greens functions of
the perfect harmonic lattice for calculations of strongly localized
ILMs, our method works generally, including the cases when ILMs have
remarkable size. To test the theory, we consider the even and odd
ILMs in a monatomic chain. In parallel, the molecular dynamics
simulations of these ILMs are carried out. The comparison of the
results of both calculations shows an excellent agreement.

\section{Self-consistency equations}
Let us start with the classical equations of the motion of atoms
in a lattice
\begin{equation}
M_{n} \ddot{u}_{n} = -\sum_{k} \sum_{n_{1} n_2...} V^{(k)}_{n
n_1...n_{k-1}} {u}_{n_1} {u} _{n_2}...{u}_{n_{k-1}}. \label{ddotU}
\end{equation}
Here ${u}_{n}$ is the Cartesian displacement of the atom $n$,
$M_{n}$ is its mass, $V^{(2)},\, V^{(3)},\ldots$ are the harmonic
and anharmonic springs; the subscripts $n$ include both the site
number and the number of the Cartesian component. We are considering
a localized solution of the equation (\ref{ddotU}) describing the
stable ILM. The corresponding displacements of atoms are of the form
\[u_n(t) = \xi_n + A_n \cos{\omega_l t} + O(\omega_l), \]
where $\omega_l$ is the frequency of the ILM which lies outside
the phonon spectrum, $|A_n|$ is the amplitude of the ILM at the site $n$,
$\xi_{n}$ is the shift of the equilibrium position of the atom $n$
(i.e. the dc-component of the ILM; this component differs from zero
due to odd anharmonicities), $O(\omega_l)$ is the sum of the higher
harmonics. The contributions of these harmonics are usually rather
small \cite{dolg,sivtak,Page,KisBick,SievPage,flach,LaiSi}. Below
they will be neglected.

In our further consideration, we shall make use of the fact that
$u_n(t+\tau))$ is also a periodic solution of the equation
(\ref{ddotU}). Taking $\tau$ to be infinitesimal, we can find small
variation of the ILM $q_n (t) = u_n(t+\tau)- u_n (t)= - A_n \tau
\sin{\omega_l t}$, which also oscillate in time with the frequency
$\omega_l$ and amplitudes $\sim A_n$ (note the phase shift $\pi/2$
of $q_n(t)$); it satisfies the equation:
\begin{equation}
M_n \ddot{q}_n = -\sum_{n'} (V^{(2)}_{n n'} + \partial^2
V_{\mathrm{anh}}/\partial u_n \partial u_{n'}) q_{n'}, \label{Eqnqn}
\end{equation}
where $V_{\mathrm{anh}}$ is the anharmonic part of the potential energy. We
consider the $\propto \sin{\omega_l t}$ terms of this equation (i.e.
we neglect higher-order harmonics of it). Taking into account that
the $\propto \sin{\omega_l t}$ term of the product
$\cos^{2n}{\omega_l t} \sin{\omega_l t}$ coincides
with the time-independent term of the product
$2 \cos^{2n}{\omega_l t} \sin^{2}{\omega_l t},$ we get
\begin{equation}
M_n \ddot{q}_n = -\sum_{n'} (V^{(2)}_{n n'} + v_{n n'}) q_{n'},
\label{Eq_q_n}
\end{equation}
where
\begin{equation}
v_{nn'} =2 \langle \sin^2{\omega_l t} \, \partial^2
V_{\mathrm{anh}}/\partial u_n \partial u_{n'} \rangle. \label{vPerturb}
\end{equation}
Here the partial derivative is taken for $u_n = \xi_n + A_n
\cos{\omega_l t}$,
\[
\langle \dots \rangle =  \frac{\omega_l}{2
\pi} \int_0^{2\pi/\omega_l}\dots dt
\]
denotes the averaging over the period $2\pi/\omega_l$.

The dc-shifts $\xi_n$ are not the independent parameters of the
problem: they are determined by the ILM amplitudes. The
corresponding relation reads \cite{HNS}
\begin{equation}
\xi_n =  \sum_{n'} g_{nn'} \langle \partial V_{\mathrm{anh}} /\partial u_{n'}
\rangle, \label{xi_n}
\end{equation}
where $g_{nn'} ={G^{(0)}_{n n'} (0)}/{\sqrt{M_n M_{n'}}},$
$G^{(0)}_{nn'} (0)$ is the static limit ($\omega=0$) of the Green's
function of the perfect lattice \cite{maradudin,economou}
\[ G^{(0)}_{nn'} (\omega) = \sum_j e_{nj} e_{n'
j}/(\omega^2-\omega_j^2), \]
$e_{nj}\propto e^{ik_j n}$ is the polarization vector of the phonon
$j$, $k_j$ is its wave vector.


Our further consideration is based on the observation that the
equation (\ref{Eq_q_n}) corresponds to the following harmonic
potential energy
\begin{equation}
V_{\mathrm{MF}}=\frac{1}{2} \sum_{nn'}\big(V^{(2)}_{nn'} + v_{nn'}\big) q_n
q_{n'}. \label{Vtilde}
\end{equation}
This potential energy, if the amplitude parameters $A_n$ are chosen
correctly, should lead to the appearance of the linear local mode,
being an infinitesimal part of the ILM. The potential $V_{\mathrm{MF}}$
constitutes a mean field for the linear mode and, therefore for the
ILM. The matrix $v_{nn'}$ gives the required change of the elastic
springs.

The potential $V_{\mathrm{MF}}$ should be determined self-consistently. This
can be done by applying the Lifshitz method \cite{maradudin} of the
local dynamics in harmonic approximation. In this method, the
amplitude parameters satisfy the relations
\begin{equation}
A_n/A_0 = G_{n0}(\omega_l)/ G_{00}(\omega_l),
\label{Self} \end{equation}
where $G(\omega)$ is the matrix of the perturbed Green's function
being equal \cite{maradudin}
\begin{equation}
G (\omega) = \big(I-G^{(0)} (\omega)\, v \big)^{-1} G^{(0)}
(\omega), \label{Green_omega}
\end{equation}
the frequency $\omega_l$ is given by the position of the pole
of $G(\omega)$ outside the phonon spectrum.
Taking into account that the amplitude parameters $A_n$ enter into
the perturbation matrix $v$, one can conclude that equations
(\ref{Self}) together with the pole condition $ \big(I-G^{(0)}
(\omega) \,v \big)^{-1} \rightarrow \infty $ constitute the
self-consistency equations.

To calculate an ILM by means of the proposed method, one should
choose the main amplitude(s) of the ILM and to fix a possible values
of few other amplitudes. Using these amplitudes one can find the
dc-shifts $\xi_{n}$ by solving the equation (\ref{xi_n}). Inserting
obtained $\xi_{n}$ into Eq. (\ref{vPerturb}) one can calculate
matrix $v$. After that one should calculate the frequency of the ILM
and the relative amplitudes, using equations (\ref{Green_omega}) and
(\ref{Self}). If the obtained  $A_n$ differ from the chosen
initially, the calculations should be repeated with newly found
amplitudes.

The presented theory can be applied for different, including 3D,
anharmonic lattices.

\section{Test for monatomic chain}

As a test of the presented theory we consider the well-known case of
ILMs in a monatomic chain, taking into account only the
nearest-neighbor interactions and the positive quartic
anharmonicity. The potential energy of vibrations in this case has
the form
\begin{equation}
V= \frac{1}{2}\sum_n \bar{u}_n^2 \left[K_{2} + \frac{1}{2}
K_{4}\bar{u}_n^2 \right], \label{Vchane}
\end{equation}
where  $\bar{u}_n= u_n - u_{n-1}$ is the difference of the shifts of
the particles $n$ and $n-1$. The top phonon frequency equals
$\omega_{\mathrm{m}} = 2\sqrt{K_2}$. In our calculations we took $K_2=100$
which corresponds to $\omega_{\mathrm{m}}=20$. The Green's functions of the
harmonic chain equal \cite{economou}
\begin{equation}
G^{(0)}_{nn'}(\omega) = [-\rho(\omega)]^{|n-n'|}/ \omega
\sqrt{\omega^2 -1}, \label{cheneGreen}
\end{equation}
where $\rho (\omega) = (\omega - \sqrt{\omega^2-1})^2 \leq 1$ (in
 Eq. (\ref{cheneGreen}) the units $\omega_{\mathrm{m}} =M=1$ are used).

The dc-shifts in this model equal zero, and the equation for the
perturbation matrix $v$ can be found strightforward:
\begin{equation}
v_{nn'} = \delta_{n,\,n'} (\gamma_{n+1} + \gamma_{n}) -
\delta_{n-1,\,n'} \gamma_n - \delta_{n+1,\,n'} \gamma_{n+1}.
\label{ChainPerturb}
\end{equation}
Here
\begin{equation}
\gamma_n = \frac{3}{4} K_4 \bar{A}_n^2  \label{digamma}
\end{equation}
is the renormalization  of the elastic spring between the atoms $n$
and $n-1$, $\bar{A}_n= A_n-A_{n-1}$. For $K_4 >0$ all $\gamma_n$ are
positive. Therefore the ILMs appear above the phonon spectrum; they
have an odd or an even symmetry.

First we study an even ILM centered at the atoms $n=0$ and $n=-1$.
In this case, the amplitude parameters satisfy the condition
$A_{n-1} =- A_{-n}$. We consider only ILMs which have the largest
amplitude on the  atoms $n=0$ and $n= -1$ and take into account the
contributions of $8$ central atoms. In this case we include into
consideration the renormalization of $7$ central springs given by
$\gamma_0, \, \gamma_{\pm 1}, \, \gamma_{\pm 2}$
and $ \gamma_{\pm 3}$ (see Eq. (\ref{digamma})). These parameters
can be determined together with the frequency and the amplitudes
self-consistently using the equations (\ref{Green_omega}) and
(\ref{Self}). The latter we treat iteratively: we start with the
amplitude ratios given in \cite{Page} and then find the frequency,
the values of the amplitudes and then their corrections from these
equations and so on. The iteration procedure converges fast.
The results of the calculations can be seen in Table 1  and in Fig.
1 (theory).

Now we consider an odd ILM centered at the $n=0$ atom. Then the
amplitude parameters of the ILMs satisfy the condition $A_n =
A_{-n}$. We restrict ourself with consideration only of ILMs having
the largest amplitude on the central atom, and take into account the
contribution of $7$ central atoms. In this case we take into account
the changes by the ILMs of  $6$ central springs given by $\gamma_0,
\, \gamma_{\pm 1}, \, \gamma_{\pm 2}$ and $\gamma_{3}$ (see Eq.
(\ref{digamma})). As in the case of even ILMs, the  springs
parameters are determined together with the frequency and the
amplitudes self-consistently using the equations (\ref{Green_omega})
and (\ref{Self}). The results of the calculations can be seen
in Fig. 1 (theory) and in Table 1.

\vspace{-0.4mm}

\section{Comparison with MD calculations}

\vspace{-0.4mm}

In the case under consideration, one can easily find the ILMs
numerically by integrating the equations of motion. In our
calculations we used the forth-order Runge-Kutta algorithm. The
results of calculations are compared with the theory. In Fig. 1 and
in Table 1  one can see that the theoretical and the MD calculations
of the ILMs frequencies and amplitudes are in very good agreement.

\begin{figure}[th]
\begin{center}
\includegraphics*[angle=-90,width=.70\textwidth]{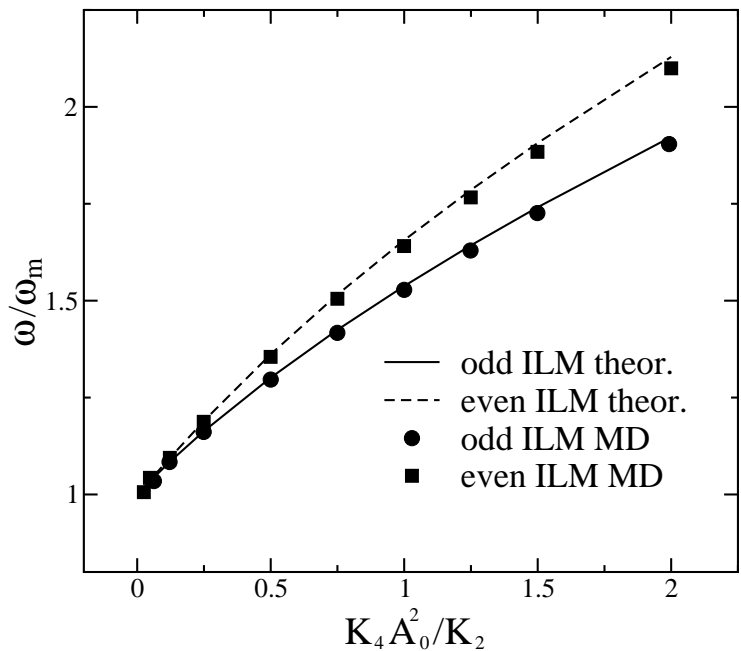}
\end{center}
\hspace*{0em} \caption{ Frequency dependences of the intrinsic
localized modes in a monatomic chain with a hard quartic
anharmonicity versus the dimensionless anharmonicity parameter
$ K_4 A_0^2 /4 K_2$; $|A_0|$ is the ILM amplitude on the central atom.}
\end{figure}

\vspace{5mm}
\begin{table}

\caption {Frequencies and amplitudes of the even and odd ILMs in a
monatomic chain with a hard quartic anharmonicity. For every $K_4$
two values of the parameters mentioned are given:  the theoretical
value (up) and the result of the MD calculation (down).}

\vspace{5mm}

\begin{center}
\begin{tabular}  {ccccccc}

\hline \hline

 \multicolumn {7}{c}{Even ILM} \\

\hline

\vspace{0.5mm} \ $ K_4$ \
 & \ $\frac{{\textstyle K}_{4}} {{\textstyle K}_{2}}A_0^2$ \
 & \ $ \frac{{\textstyle\omega}_{l}} {{\textstyle \omega}_{m}}$ \
 & \ \ \ $A_0$ \ \ \ & \ \ \ $-A_1$ \ \ \
 & \ \ \ $A_2$ \ \ \ & \ \ \ $-A_3$ \ \ \ \\

\vspace{-2.5mm} 50 & 0.1250 & 1.0963 & 0.5000& 0.3211 & 0.1537 & 0.0662 \\

 & 0.1224 & 1.0945 & 0.4947& 0.3208 & 0.1550 & 0.0680 \\

\vspace{-2.5mm}  200 & 0.5000 & 1.3617 & 0.5000& 0.1981 & 0.0419 & 0.0084 \\

 & 0.4996 & 1.3551 & 0.4998& 0.1981 & 0.0421 & 0.0083 \\

\vspace{-2.5mm}  800 & 2.0000 & 2.1289 & 0.5000& 0.1218 & 0.0084 & 0.0005 \\

\vspace{0.5mm} & 1.9999 & 2.0995 & 0.4999& 0.1219 & 0.0085 & 0.0005 \\

 \hline
 \hline

 \multicolumn {7}{c}{Odd ILM} \\

\hline

\vspace{1.mm} $K_4$ & $\frac{{\textstyle K}_{4}} {{\textstyle
K}_{2}}A_0^2$
 & $ \frac{{\textstyle\omega}_{l}} {{\textstyle \omega}_{m}}$
 & $A_0$ & $-A_1$  & $A_2$ & $-A_3$ \\

\vspace{-2.5mm}  50 & 0.1250 & 1.0862 & 0.5000& 0.4034 & 0.2266 & 0.1054 \\

 & 0.1210 & 1.0839 & 0.4920 & 0.3989 & 0.2288 & 0.1117 \\

\vspace{-2.5mm}  200 & 0.5000 & 1.3014 & 0.5000& 0.3315 & 0.0999 & 0.0225 \\

 & 0.4998 & 1.2965 & 0.4999& 0.3314 & 0.1001 & 0.0229 \\

\vspace{-2.5mm}  800 & 2.0000 & 1.9228 & 0.5000& 0.2858 & 0.0386 & 0.0031 \\

 & 1.9920 & 1.9040 & 0.4990 & 0.2726 & 0.0358 & 0.0029 \\

\hline \hline

\end{tabular}

\end{center}

\vspace{3.mm}

\end{table}

Note that the small discrepancies between the theoretical and the MD values
of the ILM  frequency $\omega_l$ increases with $\omega_l$.
The reason is that we have neglected the higher harmonics: the contribution of
these harmonics to an ILM is the larger the larger is the frequency.

\section{Conclusion}

To conclude, we have developed a theory of intrinsic localized modes
in anharmonic lattices, which allows one to make use of harmonic
approximation results and to perform calculations for a
macroscopically large lattice. We have derived the equations
describing small variations of ILMs amplitudes, which allowed us to
reduce the original nonlinear problem to the linear problem of
phonon localization on the local effective potential. The latter is
created by the ILM itself and it is determined self-consistently.
This enabled us to apply the Greens' function method  of the
perturbed local dynamics of phonons (the Lifshitz method) for the
calculations of ILMs. The theory works generally, including the
cases when ILMs have remarkable size. To test the theory we
calculated even and odd ILMs in an anharmonic monatomic chain.  We
have also carried out the MD calculations of the ILMs in the chain.
A comparison of the results, obtained by both methods (see Fig. 1
and Table 1) gives full confirmation of the proposed theory.
Although concrete calculations here are performed only for the
monatomic chain, the theory itself is general and can be applied for
different lattices, including 3D crystals with arbitrary anharmonic
potentials.

\section{Acknowledgement}

This research was supported by the Estonian
Science Foundation, Grants No.~6534 and No.~6540 and the NRG
Grant. The authors are grateful to A. J. Sievers for multiple
helpful discussions.


\end{document}